\documentstyle[12pt]{article}
\newcommand{\eq}{\begin{equation}}
\newcommand{\en}{\end{equation}}
\newcommand{\eqa}{\begin{eqnarray}}
\newcommand{\ena}{\end{eqnarray}}
\newcommand{\eqs}{\begin{displaymath}}
\newcommand{\ens}{\end{displaymath}}
\newcommand{\eqas}{\begin{eqnarray*}}
\newcommand{\enas}{\end{eqnarray*}}

\advance\oddsidemargin by -\textwidth
\advance\oddsidemargin by -1in
\evensidemargin=\oddsidemargin
\begin{document}

$\mbox{ }$
\vspace{-3cm}
\begin{flushright}
\begin{tabular}{l}
{\bf KEK-TH-650 }\\
{\bf KEK preprint 98 }\\

\end{tabular}
\end{flushright}
 
\baselineskip18pt
\vspace{1cm}
\begin{center}
\Large
{\baselineskip26pt \bf 
A Relation between \\
Commutative and Noncommutative \\
Descriptions of D-branes
}
\end{center}
\vspace{1cm}
\begin{center}
\large
{ Nobuyuki Ishibashi}
\footnote{E-mail:ishibash@theory.kek.jp 
}
\end{center}
\normalsize
\begin{center}
{\it KEK Theory Group, Tsukuba, Ibaraki 305, Japan}
\end{center}
\vspace{2cm}
\begin{center}
\normalsize
ABSTRACT
\end{center}
{\rightskip=2pc 
\leftskip=2pc 
\normalsize
In string theory, D-branes can be expressed as a configuration 
of infinitely many lower dimensional D-branes. Using this relation, 
the worldvolume theory of 
D-branes can be regarded as the worldvolume theory of the infinitely many 
lower dimensional branes. 
In the description in terms of the lower dimensional branes, 
some of the worldvolume coordinates 
become noncommutative. 
Actually this noncommutative theory can be regarded as 
noncommutative Yang-Mills theory. 
Therefore the worldvolume theory of D-branes have two equivalent
descriptions, namely the usual static gauge description using 
ordinary Yang-Mills theory and 
the noncommutative description using noncommutative Yang-Mills theory.  
It will be shown that these two descriptions correspond to two different ways 
of gauge fixing of the reparametrization invariance and its generalization. 
We will give an explicit relation between commutative gauge field and 
noncommutative gauge field in semiclassical approximation, 
when the gauge group is $U(1)$. 
\vglue 0.6cm}

\newpage
\section{Introduction}
Many physicists are interested in noncommutative geometry, 
because they expect that it captures 
some features of quantum gravity. 
It is intriguing to see that in string theory and M theory, 
which is considered to 
be the most promising model of quantum gravity, we come across 
several occasions in which space-time coordinate becomes 
noncommutative\cite{Witten:1996im}-\cite{Lizzi:1997yr}. 
In this note we would like to discuss an example in which noncommutativity of 
spacetime coordinates 
appears in string theory. The example we study here is 
the worldvolume theory of D-branes. 
As was pointed out in \cite{townsend, dhn}, 
D$p$-branes can be represented as a configuration 
of infinitely many D$(p-2r)$-branes. If such a relation hold, the 
worldvolume theory of the D$p$-branes can also be regarded as 
the worldvolume theory 
of infinitely many D$(p-2r)$-branes. 
We will show that some of the coordinates on 
the worldvolume of the D$p$-branes become noncommutative 
if one consider it as the worldvolume theory of D$(p-2r)$-branes. 
Actually the noncommutative theory we have is noncommutative Yang-Mills theory. 
Such a noncommutative description of the D$p$-branes should be 
equivalent to the usual commutative descriptions. 
We will pursue this equivalence and show that these two descriptions 
correspond to two different way to fix the reparametrization invariance 
and its generalization on the worldvolume. 
Therefore we have here an example where a noncommutative theory is equivalent 
to a commutative theory. Such an equivalence in a similar context was 
studied in a 
recent paper\cite{SW}. We will comment on the relation between our results 
and theirs. 

In section 2, we explain how D$p$-branes can be expressed as a configuration of 
infinitely many D$(p-2r)$-branes. 
In section 3, we study the worldvolume theory of the D$p$-branes regarding it 
as a configuration of the D$(p-2r)$-branes. 
Section 4 is devoted to discussions. 

This note is based on a talk presented at 
``Workshop on Noncommutative Differential 
Geometry and its Application to Physics'', Shonan-Kokusaimura, Japan, 
May31-June 4, 1999.

After I completed this note, I was informed that there are 
papers \cite{Cornalba:1999hn} \cite{Cornalba:1999ah}
whose results have considerable overlap with ours. 
Especially, in the first paper, 
they realized that the commutative and noncommutative 
descriptions of D-branes correspond to two different ways of gauge 
fixing, and the explicit relation between the commutative and 
noncommutative gauge fields, which coincides with ours were given 
in \cite{Cornalba:1999ah}. 

\section{D$p$-branes from D$(p-2r)$-branes}
In this section we will explain how D$p$-branes can be expressed as a 
configuration of infinitely many D$(p-2r)$-branes. 
For simplicity, the special case of expressing one D$p$-brane 
by D-instantons, namely $p=2r+1$ case, will be treated here. 
We will comment on more general cases 
at the end of this section. 

We study this problem in the Euclidean space ${\bf R}^D$.
\footnote{Construction of D$p$-branes from D$(p-2r)$-branes was done 
on torus in \cite{Ganor:1997zk}. 
Things discussed here are partially generalized to the 
space compactified on a torus in \cite{Kato:1999mf}.}
$D$ should be $26$ for bosonic string and $10$ for superstring. 
We will first deal with bosonic string theory in which the whole 
manipulations are simpler. 
Later we will explain how one can generalize the arguments to the 
superstring case. 
The configuration of infinitely many D-instantons can be expressed by the 
$\infty \times \infty$ hermitian matrices $X^M~(M=1,\cdots ,D)$.
The one we consider is 
\eqa
X^i
&=&
\hat{P}^i,~(i=1,\cdots ,p+1)
\nonumber
\\
X^M
&=&
0~(M=p+2,\cdots ,D),
\label{PQ-1}
\ena
where $\hat{P}^i~(i=1\cdots ,p+1)$ satisfy 
\eq
[\hat{P}^i,\hat{P}^j]=i\theta^{ij}.
\label{com}
\en 
In this note, the $(p+1)\times (p+1)$ matrix $\theta$ is assumed to 
be invertible. 

Let us show that this configuration of D-instantons is equivalent to 
a D$p$-brane. In order to show that two different configurations of 
D-branes are equivalent, one should prove that the open string theory 
corresponding to these configurations are equivalent. 

A quick way to see the equivalence is to look at the boundary states.  
The boundary state $|B\rangle$ corresponding to the configuration 
eq.(\ref{PQ-1}) can be written as follows:
\eq
|B\rangle =\mbox{TrP}e^{-i\int_0^{2\pi}d\sigma p_i\hat{P}^i}|B\rangle_{-1}.
\label{B}
\en
Here $p_i(\sigma )$ is the variable conjugate to the string coordinate 
$x^i(\sigma )$ and is equal to $\frac{1}{2\pi\alpha^\prime}\dot{x}_i$ in 
the usual flat background. $|B\rangle_{-1}$ denotes the boundary state for 
a D-instanton at the origin and satisfies $x^i(\sigma )|B\rangle_{-1}=0$. 
$|B\rangle_{-1}$ includes also the ghost part which is not relevant to the 
discussion here. 
The factor in front of $|B\rangle_{-1}$ is an analogue of the Wilson loop 
and corresponds to the background eq.(\ref{PQ-1}). 
Eq.(\ref{B}) can be rewritten by using path integral as
\eq
|B\rangle =\int [dP]
e^{\frac{i}{2}\int d\sigma P^i\partial_\sigma P^j\omega_{ij}
-i\int d\sigma p_iP^i}|B\rangle_{-1},
\label{path}
\en
where $\omega_{ij}=(\theta^{-1})_{ij}$. 

It is straightforward to perform the path integral in eq.(\ref{path}). 
Using the Fock space representation of $|B\rangle_{-1}$, the path integral is 
Gaussian. The determinant factor can be regularized in the usual way
\cite{Callan:1988wz}, and one can show that $|B\rangle$ coincides with the 
boundary state 
for a D$p$-brane with the $U(1)$ gauge field strength 
$F_{ij}=\omega_{ij}$ on the worldvolume. 
Knowing that the path integration is Gaussian, 
it is easy to confirm this fact. Indeed, one can show that the 
following identity holds:
\eqa
0
&=&
\int [dP]\frac{\delta}{\delta P^i(\sigma )}
e^{\frac{i}{2}\int d\sigma P^i\partial_\sigma P^j\omega_{ij}
-i\int d\sigma p_iP^i}|B\rangle_{-1}
\nonumber
\\
&=&
[i\omega_{ij}\partial_{\sigma}x^j-ip_i(\sigma )]|B\rangle .
\ena
Therefore $|B\rangle$ should coincide with 
$\exp [\frac{i}{2}\int d\sigma x^i\partial_\sigma x^j\omega_{ij}]|B\rangle_{p}$ 
up to normalization. Here $|B\rangle_{p}$ denotes the boundary state for 
a D$p$-brane satisfying $p_i(\sigma )|B\rangle_{p}=0$. 
Hence $|B\rangle$ is equivalent 
to the boundary state for a D$p$-brane with the $U(1)$ gauge field strength 
$F_{ij}=\omega_{ij}$ on the worldvolume. 

In the above analysis using the path integral expression eq.(\ref{path}), 
the overall normalization of the boundary state is ambiguous. 
Actually one can prove the equivalence including the normalization by 
showing that the open string theory corresponding to the 
configuration eq.(\ref{PQ-1}) is equivalent to the one corresponding to 
a D$p$-brane. We refer to \cite{ishibashi} for more details. 

It is easy to supersymmetrize the above arguments. 
In the NSR formalism the boundary state for D-instanton can be written as 
a sum of four states $|B;\pm\rangle_{-1,I}$ where $I=NS,R$ indicates the  
sector it belongs to and 
\begin{eqnarray}
& &
x^M (\sigma )|B;\pm\rangle_{-1,I}=0,
\nonumber
\\
& &
(\psi^M (\sigma )\pm i\tilde{\psi}^M (\sigma ))|B;\pm\rangle_{-1,I}=0.
\end{eqnarray}
Supersymmetric generalization of eq.(\ref{path}) can be given for each 
$|B;\pm\rangle_{-1,I}$ as 
\begin{equation}
|B;\pm\rangle_{I}=\int [dPd\chi ]
e^{\frac{1}{2}\int d\sigma 
(iP^i\partial_\sigma P^j+\chi^i\chi^j)\omega_{ij}
-\int d\sigma 
(ip_iP^i-\pi_i\chi^i)}|B;\pm \rangle_{-1,I},
\label{spath}
\end{equation}
where
$\pi^M (\sigma )=\frac{1}{2}(\psi^M(\sigma ) \mp i\tilde{\psi}^M(\sigma ))$. 
 We can show following the same arguments as above that this boundary 
state coincides with the boundary state for a D$p$-brane up to 
normalization. It is also possible to prove the equivalence of the open 
string theories \cite{super}. 

Since the arguments in this section are essentially about the variables 
$x^i,\psi^i~(i=1,\cdots ,p+1)$ on the worldsheet, 
it is quite straightforward to apply the arguments here to prove that 
a D$p$-brane can be expressed as a configuration of infinitely many 
D$(p-2r)$-branes. 
It is also easy to generalize the argument to 
the case of $N$ D$p$-branes. In such a case we should consider the 
block diagonal background
\begin{equation}
X^i=\hat{P}^i\otimes I_N,
\label{Nback}
\end{equation}
where $I_N$ is the $N\times N$ identity matrix which is an element of 
$U(N)$ Lie algebra. 
The expression of the boundary state in eq.(\ref{path}) should be modified to 
\begin{equation}
|B\rangle =\int [dP]\mbox{TrP}
e^{\frac{i}{2}\int d\sigma P^i\partial_\sigma P^j\omega_{ij}
-i\int d\sigma p_iP^i}|B\rangle_{-1},
\label{pathN}
\end{equation}
where TrP here means the trace of the path-ordered product with respect to 
the $U(N)$ indices. 
It is easy to see that this configuration corresponds to $N$ D$p$-branes 
following the same arguments as above.

\section{Worldvolume Theory}
In the previous section, we explained that the open string theory corresponding 
to the configuration of D-instantons in eq.(\ref{PQ-1}) is equivalent to 
the one for a D$p$-brane with $F=\omega$. This means that the worldvolume 
theory of one D$p$-brane can also be described as the worldvolume theory of 
D-instantons. 
In this section, we investigate the worldvolume theory from two different 
points of view, i.e. either as the worldvolume theory of one D$p$-brane or 
D-instantons.
We call them D$p$-brane picture and D-instanton picture respectively.  
We will show how these two are related with each other. 
The argument in this section 
will be done for bosonic string case for simplicity. For superstring case, 
similar results can be derived starting from the expression eq.(\ref{spath}). 

Our argument in this section can be applied to study the worldvolume theory 
of D$p$-branes by regarding it as the worldvolume theory of infinitely many 
D$(p-2r)$-branes for $2r<p+1$.  
It is also straightforward to deal with the case when the number of the 
D$p$-brane is 
more than one. We will comment on these generalizations at the end of this 
section.

\subsection{D$p$-brane Picture}
Let us start from the following expression of the boundary state for the 
D$p$-brane:
\begin{equation}
|B\rangle =\int [dP]
e^{\frac{i}{2}\int d\sigma P^i\partial_\sigma P^j\omega_{ij}
-i\int d\sigma p_iP^i}|B\rangle_{-1}.
\label{F=-o}
\end{equation}
This corresponds to a D$p$-brane longitudinal to $x^i~(i=1,\cdots ,p+1)$ 
directions with the $U(1)$ gauge field strength $F=\omega$. 
The worldvolume theory of a D$p$-brane consists of a gauge field $A_i$ and 
scalar 
fields $\phi^M~(M=p+2,\cdots ,D)$.  $\phi^M$ correspond to the shape 
of the worldvolume which can be expressed by the equation 
$x^M=\phi^M(x^1,\cdots ,x^{p+1})$. 
We are considering here the field configurations in the static gauge, 
so that the coordinates on the worldvolume are taken to be 
$x^1,\cdots ,x^{p+1}$. 
The boundary state corresponding to 
a configuration of $A_i,\phi^M$ is easily guessed to be 
\begin{equation}
|B\rangle =\int [dP]
\exp [i\int d\sigma A_i(P)\partial_\sigma P^i
-i\int d\sigma (p_iP^i+\sum_{M=p+2}^{D}p_M\phi^M(P))]|B\rangle_{-1}.
\label{Dp}
\end{equation}
Indeed this coincides with eq.(\ref{F=-o}) when $F=\omega ,\phi^M=0$. 
For small deformations $\delta A_i,\delta \phi^M$ from the background 
$F=\omega ,\phi^M=0$, 
we expect that the 
boundary state $|B\rangle$ in eq.(\ref{F=-o}) is modified by the vertex 
operator as 
\begin{eqnarray}
& &
(1+i\int d\sigma (\delta A_i(x)\partial_\sigma x^i-\delta\phi^M(x)p_M))
\int [dP]
e^{\frac{i}{2}\int d\sigma P^i\partial_\sigma P^j\omega_{ij}
-i\int d\sigma p_iP^i}|B\rangle_{-1},
\nonumber
\\
& &
\label{vertexDp}
\end{eqnarray}
which is consistent with eq.(\ref{Dp}). 
Moreover since $x^M(\sigma )|B\rangle_{-1}=0$, this 
boundary state exactly describes the emission of a closed string from the 
worldvolume $x^M=\phi^M(x^1,\cdots ,x^{p+1})$. The contribution of the 
gauge field is in the form of the Wilson loop. 
This state will be BRS invariant 
for only those $A_i,\phi^M$ satisfying the equations of motion. 
$Q_B|B\rangle =0$ implies that the path integral measure is invariant under the 
reparametrization 
$\sigma\rightarrow\sigma^\prime (\sigma )$. Imposing such a condition, one 
may be able to deduce the equations of motion after calculations similar to 
\cite{Callan:1988wz}. 

Thus in this picture, the worldvolume theory is a $U(1)$ gauge theory with 
scalars $\phi^M$. In this note, we always assume that Pauli-Villars 
regularization 
on the worldsheet is taken so that the noncommutativity because of the 
regularization 
discussed in \cite{SW} does not occur.  

\subsection{D-instanton Picture}
Now let us consider the worldvolume theory as the worldvolume theory of 
D-instantons. The boundary state eq.(\ref{F=-o}) 
corresponds to the configuration eq.(\ref{PQ-1}) of D-instantons. 
General configuration of D-instantons can be described as 
\begin{equation}
X^M=\phi^M(\hat{P})~(M=1,\cdots ,D),
\label{general}
\end{equation}
if one assumes that the operators $\hat{P}^i~(i=1,\cdots ,p)$ generate all the 
possible operators acting on the Chan-Paton indices of D-instantons. 
Here in defining the functions $\phi^M$, we need to specify the ordering of the 
operators $\hat{P}^i$, which will be given shortly.  
What will be the form of the boundary state corresponding to the configuration 
eq.(\ref{general})? A natural guess is 
\begin{equation}
|B\rangle =\int [dP]
e^{\frac{i}{2}\int d\sigma P^i\partial_\sigma P^j\omega_{ij}
-i\int d\sigma p_M\phi^M(P)}|B\rangle_{-1}.
\label{D-1}
\end{equation}
For small deformations $\delta \phi^M$ from the background eq.(\ref{PQ-1}), 
we expect that the boundary state in eq.(\ref{F=-o}) is modified as 
\begin{eqnarray}
& &
\int [dP](1-i\int d\sigma \delta\phi^M(P)p_M)
e^{\frac{i}{2}\int d\sigma P^i\partial_\sigma P^j\omega_{ij}
-i\int d\sigma p_iP^i}|B\rangle_{-1}.
\nonumber
\\
& &
\label{vertexD-1}
\end{eqnarray}
Since the vertex operators corresponding to the transverse deformations 
$\phi^M~(M=p+2,\cdots ,D)$ coincides with those in eq.(\ref{vertexDp}), 
we expect that the transverse $\phi^M$ appear in the same way as in 
eq.(\ref{Dp}). Eq.(\ref{D-1}) is unique choice satisfying this condition and 
the $D$-dimensional rotational invariance. 
The boundary state eq.(\ref{D-1}) describes 
the emission of a closed string from the hypersurface $X^M=\phi^M(P)$. 
Therefore the fields 
$\phi^M(P)$ correspond to the shape of the D-branes and $P^i$ play the role of 
the  
coordinates of the $p$-brane. 

In order to be consistent with the path integral form eq.(\ref{D-1}), 
the ordering in eq.(\ref{general}) should be chosen to be 
Weyl-ordering\cite{Lee}. To be more 
explicit, for each c-number function $f(P)$, let us define the Weyl-ordered 
function $f(\hat{P})$ to be 
\begin{equation}
f(\hat{P})=\int d^{p+1}ke^{ik_i\hat{P}^i}\tilde{f}(k),
\label{operator}
\end{equation}
where 
\begin{equation}
\tilde{f}(k)=\int \frac{d^{p+1}P}{(2\pi )^{p+1}}e^{-ik_iP^i}f(P).
\label{cnumber}
\end{equation}
Then $\phi^M(\hat{P})$ on the right hand side of eq.(\ref{general})  
should be understood to be the Weyl-ordered function corresponding to 
the c-number function $\phi^M(P)$ in eq.(\ref{D-1}).  

The action and other physical quantities in the worldvolume theory of 
D-instantons 
are written as a trace of a function of the Weyl-ordered operators in 
eq.(\ref{operator}). 
However it is more convenient for us to rewrite everything in terms of 
the c-number 
functions $\phi^M(P)$ in eq.(\ref{cnumber}). 
We can do so by using the following formula\cite{Aoki:1999vr}:
\begin{equation}
Tr(f_1(\hat{P})f_2(\hat{P})\cdots f_n(\hat{P}))
=
\int \frac{d^{p+1}P}{(2\pi )^{(p+1)/2}\sqrt{|\mbox{det}\theta |}}
f_1(P)*f_2(P)\cdots f_n(P).
\end{equation}
Here the $*$-product is defined as
\begin{equation}
f(P)*g(P)=e^{\frac{i}{2}\theta^{ij}\frac{\partial}{\partial\xi^i}
\frac{\partial}{\partial\zeta^j}}
f(P+\xi )g(P+\zeta )|_{\xi =\zeta =0}.
\end{equation}
Hence, a trace of a function of Weyl-ordered operators can be rewritten 
in terms of 
the corresponding c-number functions by replacing product of operators by 
$*$-product of the corresponding c-number functions and trace by integral. 

Thus if one regards the worldvolume theory as a theory of D-instantons, 
the description 
should be noncommutative. $P^i$ can be considered as the coordinates 
on the $p$-brane and they are noncommutative under the $*$-product reflecting 
the 
commutation relation eq.(\ref{com}). Now let us discuss what kind of theory this 
noncommutative field theory is. The Lagrangian of the worldvolume theory of 
D-instantons 
can be written in terms of the commutators of the matrices $\phi^M(\hat{P})$. 
Since we started from the background in eq.(\ref{PQ-1}), 
let us express $\phi^i~(i=1,\cdots ,p+1)$ in the form of the background and the 
fluctuations around it:
\begin{equation}
\phi^i(\hat{P})=\hat{P}^i+\theta^{ij}a_j(\hat{P}).
\label{a_i}
\end{equation}
The c-number expression corresponding to the commutators of $\phi^i(\hat{P})$ 
are 
easily calculated to be 
\begin{equation}
[\phi^i(\hat{P}),\phi^j(\hat{P})]
\rightarrow
i\theta^{ij}-i(\theta f\theta )^{ij},
\end{equation}
where
\begin{equation}
f_{ij}=\partial_ia_j(P)-\partial_ja_i(P)-ia_i*a_j(P)+ia_j*a_i(P).
\end{equation}
$f_{ij}$ can be considered as the field strength of a noncommutative 
Yang-Mills field $a_i$. 
$\phi^i(\hat{P})$ essentially corresponds to the covariant derivative 
$\partial +ia$. 
Thus the commutators of $\phi^i$ with other fields give the covariant 
derivative of these fields. 
Other commutators are interpreted as the gauge covariant commutators 
in the noncommutative 
Yang-Mills theory. Since the Lagrangian is written in terms of these 
commutators,  
the noncommutative theory we have here can be considered as noncommutative 
Yang-Mills theory
\cite{Connes}. 
The gauge invariance of the theory stems from the transformation 
\begin{equation}
\delta \phi^M(\hat{P})=i[\epsilon ,\phi^M(\hat{P})].
\end{equation}
As a theory of D-instantons this is the $U(\infty )$ transformation under 
which the theory should 
is invariant. In the c-number formulation 
such a transformation corresponds to the coordinate transformation 
\begin{equation}
\delta P=\theta^{ij}\partial_j\epsilon (P).
\end{equation}
This is the coordinate transformation which preserves $\omega =\theta^{-1}$. 
If one regards $\omega$ as a symplectic form, such transformations are the 
canonical 
transformations. 
The invariance under the canonical transformation will be discussed in the 
next subsection. 

\subsection{Relation between the Two Pictures}
In the previous subsections we obtain two points of view about 
the worldvolume theory. 
Since they are supposed to describe the same thing, there should be 
correspondence between the 
two. In the D$p$-brane picture, the worldvolume fields are the gauge field 
$A_i~(i=1,\cdots ,p+1)$ 
and scalars $\phi^M~(M=p+2,\cdots ,D)$, where the coordinates on 
the worldvolume is 
taken to be $x^i~(i=1,\cdots ,p+1)$. 
On the other hand, the worldvolume fields in the D-instanton picture are 
$\phi^M(P)~(M=1,\cdots ,D)$ and $P^i$ are the coordinates on the worldvolume. 

As we noticed in the previous section, the fields  $\phi^M~(M=p+2,\cdots ,D)$ 
common to both 
correspond to each other. Therefore we should find how the fields $A_i$ and 
$\phi^i$ are related. 
Let us first consider small deformations $\delta A_i,\delta \phi^i$ 
from the background eq.(\ref{PQ-1}). 
From eq.(\ref{vertexDp}), one can see that $\delta A_i$ changes 
the boundary state 
as
\begin{equation}
\delta |B\rangle =
i\int [dP]\int d\sigma \delta A_i(P)\partial_\sigma P^i
e^{\frac{i}{2}\int d\sigma P^i\partial_\sigma P^j\omega_{ij}
-i\int d\sigma p_iP^i}|B\rangle_{-1},
\label{deltaA}
\end{equation}
which should be compared with the variation corresponding to $\delta \phi^i$ 
from eq.(\ref{vertexD-1}):
\begin{equation}
\delta |B\rangle =
-i\int [dP]\int d\sigma \delta \phi^i(P)p_i
e^{\frac{i}{2}\int d\sigma P^i\partial_\sigma P^j\omega_{ij}
-i\int d\sigma p_iP^i}|B\rangle_{-1}.
\label{deltaphi}
\end{equation}
The relation between these two variations can be derived from the following 
identity
\begin{eqnarray}
0
&=&
\int [dP]\frac{\delta}{\delta P^i(\sigma )}
e^{\frac{i}{2}\int d\sigma P^i\partial_\sigma P^j\omega_{ij}
-i\int d\sigma p_iP^i}|B\rangle_{-1}
\nonumber
\\
&=&
\int [dP][i\omega_{ij}\partial_\sigma P^j-ip_i]
e^{\frac{i}{2}\int d\sigma P^i\partial_\sigma P^j\omega_{ij}
-i\int d\sigma p_iP^i}|B\rangle_{-1},
\end{eqnarray}
which implies that $\delta |B\rangle$ in eqs.(\ref{deltaA})(\ref{deltaphi}) 
coincide with each other when $\delta A_i=\omega_{ij}\delta\phi^j$. 
Such a relation was given in \cite{miao}. 

In order to see the relation between the two pictures for 
finite deformations, the most 
convenient way is to consider a description involving fields $A_i$ and 
$\phi^M~(M=1,\cdots ,D)$. From eqs.(\ref{Dp})(\ref{D-1}), 
the boundary state involving all these fields should be 
\begin{equation}
|B\rangle =\int [dP]
e^{i\int d\sigma A_i(P)\partial_\sigma P^i
-i\int d\sigma p_M\phi^M(P)}|B\rangle_{-1}.
\label{all}
\end{equation}
However there are two many fields in such a description and there should be 
symmetry to reduce the number of them. The boundary state eq.(\ref{all}) is 
invariant under gauge transformation $\delta A_i=\partial_i\lambda$. 
Moreover, in the D$p$-brane picture, considering 
$A_i$ and $\phi^M$ just means considering the theory before the gauge fixing 
of reparametrization invariance. 
Therefore the theory should have reparametrization invariance. 

Indeed we can argue that the boundary state eq.(\ref{all}) is invariant 
under the 
following transformation
\begin{eqnarray}
& &
\delta A_i(P)
=
-\epsilon^j(P)F_{ji}(P),
\nonumber
\\
& &
\delta \phi^M(P)
=
-\epsilon^i(P)\partial_i\phi^M(P),
\end{eqnarray}
because the variation is proportional to a sum of the equations of motion 
for $P^i$. 
This transformation coincides with the coordinate transformation up to 
field-dependent 
gauge transformation because
\begin{equation}
\delta A_i(P)=
-\epsilon^j\partial_jA_i-\partial_i\epsilon^jA_j+\partial_i(\epsilon^jA_j).
\end{equation}
Since the boundary state eq.(\ref{all}) is gauge invariant, it is 
reparametrization 
invariant. 

Therefore the description involving $A_i$ and $\phi^M~(M=1,\cdots ,D)$ 
is invariant under 
the reparametrization on the worldvolume. The D$p$-brane picture obviously 
corresponds to 
the static gauge. On the other hand, one can see from eq.(\ref{D-1}) 
that the D-instanton 
picture apparently corresponds to the gauge $F_{ij}=\omega_{ij}$. 
We are not sure if 
such a gauge can be taken for arbitrary configuration of the gauge field $F$, 
but 
at least when we are thinking about the fluctuation from the background 
$F=\omega$ 
perturbatively, it seems all right. 
Such a gauge does not fix the whole 
reparametrization invariance on the worldvolume. 
The residual invariance consists of the 
coordinate transformation preserving $\omega_{ij}$, 
i.e. the canonical transformation. 
We saw such an invariance as the c-number 
counterpart of the $U(\infty )$ invariance in the previous subsection. 

Since the difference is from the gauge choice, 
we can give the explicit relation between the static gauge variables 
$A_i(x),\phi^M_{st}(x)~(M=p+2,\cdots ,D)$ 
and the variables $\phi^M_{nc}(P)~(M=1,\cdots ,D)$
\footnote{Here the subscript $st$ and $nc$ are for distinguishing 
$\phi^M$ in different 
gauges.}
 in the noncommutative description at least classically,
i.e. for small $\theta $,:
\begin{eqnarray}
& &
\phi^M_{nc}(P)
=
\phi^M_{st}(\phi^1_{nc}(P),\cdots ,\phi^{p+1}_{nc}(P))
~(M=p+2,\cdots ,D),
\\
& &
\omega_{ij}
=
F_{kl}(\phi^1_{nc}(P),\cdots ,\phi^{p+1}_{nc}(P))
\frac{\partial \phi^k_{nc}}{\partial P^i}
\frac{\partial \phi^l_{nc}}{\partial P^j}.
\label{comnc}
\end{eqnarray}
Eq.(\ref{comnc}) can be rewritten in terms of the noncommutative Yang-Mills 
field $a_i$ using eq.(\ref{a_i}) as
\begin{equation}
(\theta^{-1})_{ij}=(M^{T}F(P+\theta a) M)_{ij},
\end{equation}
where
\begin{equation}
M^i_{~j}=\delta^i_{~j}+\theta^{ik}\partial_ja_k(P).
\end{equation}
This gives an explicit relation between commutative gauge field $A_i(x)$ and 
noncommutative gauge field $a_i(P)$. 

Now let us comment on two generalizations of our results in this section. 
First one is to consider more than one D$p$-branes.  
Starting from the background eq.(\ref{Nback}), we can follow the arguments of 
$N=1$ case. This time all the fields $A_i$ and $\phi^M$ are in the adjoint 
representation of $U(N)$. 
We should put TrP in front of the right hand sides of 
eqs.(\ref{F=-o}),(\ref{Dp}),(\ref{D-1}),(\ref{all}) and then 
we can follow the same arguments as $N=1$ case. 
The reparametrization invariance of the boundary state in this case 
can be derived as follows. 
The path-ordered trace version of eq.(\ref{all}) can be rewritten by introducing 
fermions $\psi$ in the fundamental representation $U(N)$ as 
\begin{equation}
|B\rangle =\int [dPd\psi ]
\exp [\int d\sigma \psi^\dagger\partial_\sigma\psi 
-i\int d\sigma A^a_i(P)\partial_\sigma P^i\psi^\dagger t^a\psi 
-i\int d\sigma p_M\phi^{M,a}(P)\psi^\dagger t^a\psi ]|B\rangle_{-1}.
\end{equation}
Here $t^a$ are the generators of $U(N)$ in the fundamental representation. 
In this form, we can see that the boundary state is invariant under the 
transformation 
\begin{eqnarray}
& &
\delta A_i(P)
=
-\epsilon^j(P)F_{ji}(P),
\nonumber
\\
& &
\delta \phi^M(P)
=
-\epsilon^i(P)D_i\phi^M(P),
\end{eqnarray}
because the variation is proportional to the equations of motion for $P^i,\psi$. 
This transformation is again equivalent to the coordinate transformation up to 
field-dependent gauge transformation. Moreover we can argue that 
the boundary state 
is invariant under 
\begin{eqnarray}
& &
\delta A_i(P)
=
-\epsilon^j(P)t^aF_{ji}(P),
\nonumber
\\
& &
\delta \phi^M(P)
=
-\epsilon^i(P)t^a\partial_i\phi^M(P),
\end{eqnarray}
because the variation is proportional to 
$\lim_{\sigma^\prime\rightarrow\sigma }
\psi^\dagger t^a\psi (\sigma^\prime )\times 
(\mbox{equations~of~motion})(\sigma )$. 
This transformation can be considered to be a nonabelian generalization of 
coordinate transformation up to field-dependent gauge transformation. 
Fixing such invariance by taking static gauge or $F=\omega$ and 
we get commutative or noncommutative descriptions respectively. 

Secondly it is also possible to study the worldvolume theory of D$p$-branes 
regarding it as the worldvolume theory of infinitely many D$(p-2r)$-branes. 
We obtain a description in which some of the coordinates on the worldvolume 
become noncommutative.

\section{Discussions}
In this note, we have shown that D$p$-branes with constant field strength 
$F_{ij}$ can be represented as a configuration of infinitely many 
D$(p-2r)$-branes. 
The worldvolume theory of the D$p$-branes can be analysed by regarding it as the 
worldvolume theory of the D$(p-2r)$-branes and we obtain a noncommutative 
description 
of the worldvolume theory. The system we studied here is gauge equivalent to the 
one studied in \cite{SW}. In that paper, D-branes in constant $B_{ij}$ 
background 
was considered and  
the authors get commutative 
and noncommutative descriptions of the worldvolume theory depending on the 
regularization. 
Moreover they propose descriptions with continuously varying $\theta$ which 
connect the commutative and noncommutative descriptions. 
Actually we can realize such descriptions also in our formalism by 
considering our system 
in constant $B_{ij}$ background \cite{iikk}. 
Therefore we suspect that our noncommutative description is equivalent to 
a choice of 
$\theta$ in \cite{SW}. 
Since we have an explicit relation between the commutative and 
noncommutative descriptions which is valid for small $\theta$, 
it may be interesting to compare our relation and theirs. 
Here we just comment on one crucial difference. 
In \cite{SW}, the gauge transformations of the commutative 
and noncommutative descriptions are related 
but in the relation we obtained the gauge invariance of 
noncommutative theory is the residual 
reparametrization invariance which is not related to the 
commutative gauge invariance. 
Therefore the relation we obtained in eq.(\ref{comnc}) is 
for gauge invariant quantities. 

Since we have an example in which there is a relation between 
commutative and noncommutative 
theory, it may be generalized and be used in studying other 
noncommutative theories. 
One example is the noncommutative geometric formulation of 
open string field theory\cite{Witten:1986cc}. 
We think that the relation we studied in this note may be 
relevant in revealing symmetries 
hidden in the string field theory. We hope that we can come 
back to this problem in the future. 

\section*{Acknowledgements}
We would like to thank the organizers of the workshop for the wonderful 
workshop. 
I am grateful to H. Aoki, S. Iso, H. Kawai, Y. Kitazawa and T. Tada 
for collaborations and K. Okuyama for discussions. 
This work was supported by the Grant-in-Aid for Scientific Research from the 
Ministry of Education, Science and Culture of Japan. 

\newpage
%
\end{document}